\documentclass[12pt]{article}
\usepackage{amssymb}
\renewcommand{\theequation}{\arabic{section}.\arabic{equation}}

\begin{document}



\def\a{\alpha}
\def\b{\beta}
\def\d{\delta}
\def\e{\epsilon}
\def\g{\gamma}
\def\h{\mathfrak{h}}
\def\k{\kappa}
\def\l{\lambda}
\def\o{\omega}
\def\p{\wp}
\def\r{\rho}
\def\t{\tau}
\def\s{\sigma}
\def\z{\zeta}
\def\x{\xi}
\def\V={{{\bf\rm{V}}}}
 \def\A{{\cal{A}}}
 \def\B{{\cal{B}}}
 \def\C{{\cal{C}}}
 \def\D{{\cal{D}}}
\def\G{\Gamma}
\def\K{{\cal{K}}}
\def\O{\Omega}
\def\R{\bar{R}}
\def\T{{\cal{T}}}
\def\L{\Lambda}
\def\f{E_{\tau,\eta}(sl_2)}
\def\E{E_{\tau,\eta}(sl_n)}
\def\Zb{\mathbb{Z}}
\def\Cb{\mathbb{C}}

\def\R{\overline{R}}

\def\beq{\begin{equation}}
\def\eeq{\end{equation}}
\def\bea{\begin{eqnarray}}
\def\eea{\end{eqnarray}}
\def\ba{\begin{array}}
\def\ea{\end{array}}
\def\no{\nonumber}
\def\le{\langle}
\def\re{\rangle}
\def\lt{\left}
\def\rt{\right}

\newtheorem{Theorem}{Theorem}
\newtheorem{Definition}{Definition}
\newtheorem{Proposition}{Proposition}
\newtheorem{Lemma}{Lemma}
\newtheorem{Corollary}{Corollary}
\newcommand{\proof}[1]{{\bf Proof. }
        #1\begin{flushright}$\Box$\end{flushright}}

\baselineskip=20pt

\newfont{\elevenmib}{cmmib10 scaled\magstep1}
\newcommand{\preprint}{
   \begin{flushleft}
   \end{flushleft}\vspace{-1.3cm}
   \begin{flushright}\normalsize
   \end{flushright}}
\newcommand{\Title}[1]{{\baselineskip=26pt
   \begin{center} \Large \bf #1 \\ \ \\ \end{center}}}
\newcommand{\Author}{\begin{center}
   \large \bf
Junpeng Cao${}^{a}$,~Wen-Li Yang${}^b\footnote{Corresponding author:
wlyang@nwu.edu.cn}$,
 ~ Kangjie Shi${}^b$ and~Yupeng Wang${}^a\footnote{Corresponding author: yupeng@iphy.ac.cn}$
 \end{center}}
\newcommand{\Address}{\begin{center}

     ${}^a$Beijing National Laboratory for Condensed Matter
           Physics, Institute of Physics, Chinese Academy of Sciences, Beijing
           100190, China\\
     ${}^b$Institute of Modern Physics, Northwest University,
     Xian 710069, China
   \end{center}}
\newcommand{\Accepted}[1]{\begin{center}
   {\large \sf #1}\\ \vspace{1mm}{\small \sf Accepted for Publication}
   \end{center}}

\preprint
\thispagestyle{empty}
\bigskip\bigskip\bigskip

\Title{Off-diagonal Bethe ansatz solution of the XXX spin-chain with arbitrary boundary conditions}
\Author

\Address
\vspace{1cm}

\begin{abstract}
Employing the off-diagonal Bethe ansatz method  proposed recently by
the present authors, we exactly diagonalize the $XXX$ spin chain
with arbitrary boundary fields. By constructing a functional
relation between the eigenvalues of the transfer matrix and the
quantum determinant, the associated $T-Q$ relation and  the Bethe
ansatz equations are derived.

\vspace{1truecm} \noindent {\it PACS:} 75.10.Pq, 03.65.Vf, 71.10.Pm

\end{abstract}
\newpage
\section{Introduction}
\label{intro} \setcounter{equation}{0}

Our understanding of quantum phase transitions and critical
phenomena has been greatly enhanced by the study of exactly solvable
models (or quantum integrable systems) \cite{Bax82}. Such exact
results have provided valuable insight into the important
universality classes of quantum physical systems ranging from modern
condensed matter physics \cite{Kas98} to string and super-symmetric
Yang-Mills theories \cite{Dol03}. Since Yang and Baxter's pioneering
works \cite{yang2,bax1,Bax82}, the quantum Yang-Baxter equation
(QYBE), which define the underlying algebraic structure,  has become
a cornerstone for constructing and solving the integrable models.
There are several well-known methods for deriving the Bethe ansatz
(BA) solution of integrable models: the coordinate BA
\cite{Bet31,Bax82,Alc87,Bel13,Cra12}, the T-Q approach \cite{Bax82,Yan06}, the
algebraic BA \cite{Skl78,Tak79,Kor93}, the analytic BA \cite{Res83},
the functional BA \cite{Skl92} and others
\cite{And84,Baz89,Nep04,Cao03,Yan04,Gie05,Mel05,Doi06,Bas07,Gal08,Fra11,Nie09,Gra10,Nic12}.

Generally speaking, there are two classes of integrable models. One
possesses $U(1)$ symmetry and the other does not. Three well known
examples without $U(1)$ symmetry are the $XYZ$ spin chain
\cite{bax1,Tak79}, the $XXZ$ spin chain with antiperiodic boundary
condition \cite{Yun95,Bat95,Gal08,Fra11,Nie09,Nic12} and the ones
with unparallel boundary fields
\cite{Nep04,Cao03,Yan04,Gie05,Bas07,Gal08,Fra11,Nie09,Gra10,Nic12}.
It has been proven that most of the conventional Bethe ansatz
methods can successfully diagonalize the integrable models with
$U(1)$ symmetry. However, for those without $U(1)$ symmetry, only
some very special cases such as the $XYZ$ spin chain with even site
number \cite{bax1,Tak79} and the $XXZ$ spin chain with constrained
unparallel boundary fields \cite{Cao03,Yan04,Yan07,Gie05-1} can be
dealt with due to the existence of a proper ``local vacuum state" in
these special cases. The main obstacle applying the algebraic Bethe
ansatz and Baxter's method to general integrable models without
$U(1)$ symmetry lies in the absence of such a ``local vacuum". A
promising method for approaching such kind of problems is Sklyanin's
separation of variables method \cite{Skl92} which has been recently
applied to some integrable models \cite{Fra11,Nie09,Gra10,Nic12}.
However, before the very recent work\cite{Cao13}, a systematic
method was still absent to derive the Bethe ansatz equations for
integrable models without $U(1)$ symmetry, which are crucial for
studying the physical properties in the thermodynamic limit.

As for integrable models without $U(1)$ symmetry, some off-diagonal
elements of monodromy matrix enter into expression of the transfer
matrix. This breaks down the usual $U(1)$ symmetry. Very recently,
we have proposed a  method \cite{Cao13} for dealing with the
integrable models without $U(1)$ symmetry. The central idea of the
method is to construct the functional relations between eigenvalues
$\Lambda(\lambda)$ of the transfer matrix (the trace of the
monodromy matrix) and its quantum determinant $\Delta_q(\lambda)$,
i.e.,
$\Lambda(\theta_j)\Lambda(\theta_j-\eta)\sim\Delta_q(\theta_j)$ (see
below (\ref{Main})) based on the zero points of the product of
off-diagonal elements of monodromy matrix $B(u)B(u-\eta)=0$. Since
the trace and the determinant are two basic quantities of a matrix
which are independent of the representation basis, this method could
overcome the obstacle of absence of a reference state which is
crucial in the conventional Bethe ansatz methods.

Our primary motivation for this work comes from the long standing
problem of solving the open spin-$\frac{1}{2}$ XXX spin chain with
unparallel boundary fields, defined by the Hamiltonian
\cite{Skl88,Veg93}
\begin{eqnarray}
H=\sum_{j=1}^{N-1}{\vec \sigma}_{j} {\vec \sigma}_{j+1}+h_N
\sigma_{N}^{z} + h_1^x \sigma_{1}^{x}+h_1^z \sigma_{1}^{z}.
\label{ohami}
\end{eqnarray}
$N$ is the site number of the system and $\sigma_{j}^{\alpha}$
$(\alpha=x, y, z)$ is the Pauli matrix on the site $j$ along the
$\alpha$ direction. The parameters $h_N$, $h^x_1$ and $h^z_1$ are related to
boundary fields. Solving this problem for generic values of these three parameters
is a crucial step in formulating the thermodynamics of the spin chain, due to the
fact that this problem has important applications in condensed matter physics and statistical
mechanics. In this paper, we shall use the method developed in \cite{Cao13} to solve the eigenvalue problem of the
above Hamiltonian with generic $h_N$, $h^x_1$ and $h^z_1$.

The paper is organized as follows.  Section 2 serves as an
introduction our notation and some basic ingredients. We briefly
describe the inhomogeneous open XXX chain with non-diagonal boundary
terms.  In Section 3, we derive the exchange relations among the
matrix entries of the monodromy matrix algebras (or the Yang-Baxter
algebras). In section 4 after obtaining some properties of the
eigenvalue as a function of spectrum parameter $u$, we derive the
very relation between the eigenvalue and the quantum determinant of
the double-row monodromy matrix. This allows us to construct a
generalized $T-Q$ relation type solution of eigenvalue. In section
5, we consider the homogeneous limit of the results of the previous
section and give the energy spectrum of the Hamiltonian of the open
spin-$\frac{1}{2}$ XXX spin chain with unparallel boundary fields.
In section 6, we summarize our results and give some discussions.
Some detailed technical proof is given in Appendix A.


\section{ Transfer matrix}
\label{XXZ} \setcounter{equation}{0}

Throughout, ${\rm\bf V}$ denotes a two-dimensional linear space. The well-known rational
six-vertex model R-matrix $R(u)\in {\rm End}({\rm\bf V}\otimes {\rm\bf V})$
is given by \bea
R(u)=\lt(\begin{array}{llll}u+1&&&\\&u&1&\\
&1&u&\\&&&u+1\end{array}\rt).
\label{r-matrix}\eea Here $u$ is the spectral parameter. Without losing the generality we have
set the so-called bulk anisotropy parameter (or crossing parameter) $\eta=1$. The R-matrix satisfies
the quantum Yang-Baxter equation (QYBE)
\bea
R_{12}(u_1-u_2)R_{13}(u_1-u_3)R_{23}(u_2-u_3)
=R_{23}(u_2-u_3)R_{13}(u_1-u_3)R_{12}(u_1-u_2),\label{QYB}\eea
and the properties,
\bea &&\hspace{-1.5cm}\mbox{ Initial
condition}:\,R_{12}(0)= P_{12},\label{Int-R}\\
&&\hspace{-1.5cm}\mbox{ Unitarity
relation}:\,R_{12}(u)R_{21}(-u)= -\xi(u)\,{\rm id},
\quad \xi(u)=(u+1)(u-1),\label{Unitarity}\\
&&\hspace{-1.5cm}\mbox{ Crossing
relation}:\,R_{12}(u)=V_1R_{12}^{t_2}(-u-1)V_1,\quad
V=-i\s^y,
\label{crosing-unitarity}\\
&&\hspace{-1.5cm}\mbox{ PT-symmetry}:\,R_{12}(u)=R_{21}(u)=R^{t_1\,t_2}_{12}(u),\label{PT}\\
&&\hspace{-1.5cm}\mbox{ Antisymmetry}:\,R_{12}(-1)=-(1-P)=-2P^{(-)}.\label{Ant}
\eea
Here $R_{21}(u)=P_{12}R_{12}(u)P_{12}$ with $P_{12}$ being
the usual permutation operator and $t_i$ denotes transposition
in the $i$-th space. Here and below we adopt the standard
notations: for any matrix $A\in {\rm End}({\rm\bf V})$, $A_j$ is an
embedding operator in the tensor space ${\rm\bf V}\otimes
{\rm\bf V}\otimes\cdots$, which acts as $A$ on the $j$-th space and as
identity on the other factor spaces; $R_{ij}(u)$ is an embedding
operator of R-matrix in the tensor space, which acts as identity
on the factor spaces except for the $i$-th and $j$-th ones.

One introduces the ``row-to-row"  (or one-row ) monodromy matrices
$T(u)$ and $\hat{T}(u)$, which is an $2\times 2$ matrix with elements being
operators acting on ${\rm\bf V}^{\otimes N}$,
\bea
T_0(u)&=&R_{0N}(u-\theta_N)R_{0\,N-1}(u-\theta_{N-1})\cdots
R_{01}(u-\theta_1),\label{Mon-V-1}\\
\hat{T}_0(u)&=&R_{01}(u+\theta_1)R_{02}(u+\theta_{2})\cdots
R_{0N}(u+\theta_N).\label{Mon-V-2}
\eea Here $\{\theta_j|j=1,\cdots,N\}$ are
arbitrary free complex parameters which are usually called
inhomogeneous parameters.

Integrable open chain can be constructed as follows \cite{Skl88}.
Let us introduce a pair of K-matrices $K^-(u)$ and $K^+(u)$. The
former satisfies the reflection equation (RE)
 \bea &&R_{12}(u_1-u_2)K^-_1(u_1)R_{21}(u_1+u_2)K^-_2(u_2)\no\\
 &&~~~~~~=
K^-_2(u_2)R_{12}(u_1+u_2)K^-_1(u_1)R_{21}(u_1-u_2),\label{RE-V}\eea
and the latter  satisfies the dual RE \bea
&&R_{12}(u_2-u_1)K^+_1(u_1)R_{21}(-u_1-u_2-2)K^+_2(u_2)\no\\
&&~~~~~~= K^+_2(u_2)R_{12}(-u_1-u_2-2)K^+_1(u_1)R_{21}(u_2-u_1).
\label{DRE-V}\eea For open spin-chains, other than the standard
``row-to-row" monodromy matrix $T(u)$ (\ref{Mon-V-1}), one needs to
consider  the
 double-row monodromy matrix $\mathbb{T}(u)$
\bea
  \mathbb{T}(u)=T(u)K^-(u)\hat{T}(u).
  \label{Mon-V-0}
\eea Then the double-row transfer matrix of the XXX chain with open
boundary (or the open XXX chain) is given by \bea
\t(u)=tr(K^+(u)\mathbb{T}(u)).\label{trans}\eea The QYBE (\ref{QYB})
and (dual) REs (\ref{RE-V}) and (\ref{DRE-V}) lead to the fact that
the transfer matrices with different spectral parameters commute
with each other \cite{Skl88}: $[\t(u),\t(v)]=0$. Then $\tau(u)$
serves as the generating functional of the conserved quantities of
the corresponding system, which ensures the integrability of the
open XXX chain.

In this paper in order to study the system described by the Hamiltonian (\ref{ohami}),
we consider the K-matrix $K^{-}(u)$ which is a diagonal
solution to the RE (\ref{RE-V}) associated the six-vertex
model R-matrix  \cite{Veg93,Gho94}
\begin{eqnarray}
K^-(u)=\left(\begin{array}{cc}
p+u & 0\\
0 & p-u
\end{array}\right). \label{K-}
\end{eqnarray}
At the same time, we introduce  the corresponding {\it dual\/}
K-matrix $K^+(u)$ which is a generic solution to the dual
reflection equation (\ref{DRE-V})
\begin{eqnarray}
K^+(u)=\left(\begin{array}{cc}
q+u+1 & \xi(u+1)\\
\xi(u+1) & q-u-1
\end{array}\right). \label{K+}
\end{eqnarray}
The first order derivative of logarithm of the
transfer matrix $\tau(u)$ (\ref{trans}) with the K-matrices $K^{\pm}$ given
by (\ref{K-}) and (\ref{K+}) yields   the Hamiltonian (\ref{ohami})
\begin{eqnarray}
&&H=\frac{\partial \ln \tau(\lambda)}{\partial
\lambda}|_{\lambda=0,\theta_j=0}-N\nonumber \\
&&\quad =2\sum_{j=1}^{N-1}P_{j,j+1}
+\frac{{K_{N}^-}^\prime(0)}{K^-_N(0)}+2\frac{{K^+_1}^\prime(0)}{K^+_1(0)}-N \nonumber\\
&&\quad =\sum_{j=1}^{N-1}{\vec \sigma}_{j} {\vec
\sigma}_{j+1}+\frac{1}{p} \sigma_{N}^{z} +\frac{1}{q}(
\sigma_{1}^{z}+\xi \sigma_{1}^{x}). \label{oh}
\end{eqnarray}
Therefore, $h_N=1/p$, $h_1^x=\xi/q$ and $h_1^z=1/q$.


\section{ Exchange relations of the monodromy matrice}
\label{T-QR} \setcounter{equation}{0}

The Yang-Baxter algebra is a corner stone of the QISM and it has
been successfully explored in order to construct integrable systems
and to get their exact solutions.

From the quantum Yang-Baxter equation (\ref{QYB}), one may derive the following ``RLL" relations
among the one-row monodromy matrices $T(u)$ and $\hat{T}(u)$
\bea
&&R_{12}(u-v)T_1(u)T_{2}(v)=T_{2}(v)T_1(u)R_{12}(u-v),\label{RLL-1}\\
&&R_{12}(v-u)\hat{T}_2(v)\hat{T}_{1}(u)=\hat{T}_{1}(u)\hat{T}_2(v)R_{12}(v-u),\label{RLL-2}\\
&&\hat{T}_2(v)R_{12}(u+v)T_1(u)=T_1(u)R_{12}(u+v)\hat{T}(v).\label{RLL-3}
\eea
The crossing relation of the R-matrix (\ref{crosing-unitarity}) allows us to derive the following relation
between $T(u)$ and $\hat{T}(u)$
\bea
 \hat{T}_0(u)=(-1)^{N-1}V_0\,T^{t_0}(-u-1)\,V_0.\label{Crossing-TT}
\eea Let us decompose the monodromy matrices $T(u)$ and $\hat{T}(u)$
in terms of its components \bea
T(u)=\left(\begin{array}{cc} \alpha(u) & \beta(u) \\
\gamma(u)& \delta(u)
  \end{array}\right),\quad\quad
\hat{T}(u)=(-1)^{N-1}\left(\begin{array}{cc}
\delta(-u-1)&-\beta(-u-1) \\
-\gamma(-u-1)&\alpha(-u-1)
\end{array}\right). \label{Decomp}
\eea One may  find the commutation relations among $\a(u)$, $\b(u)$,
$\g(u)$ and $\d(u)$. Here we give the relevant ones for our
purpose,
\bea
&&\beta(\lambda)\beta(\mu)=\beta(\mu)\beta(\lambda),\label{Exch-1}\\
&&\beta(\lambda)\gamma(\mu)=\gamma(\mu)\beta(\lambda)+\frac{1}{\lambda-\mu}[\delta(\mu)\alpha(\lambda)-\delta(\lambda)\alpha(\mu)],
\label{Exch-2}\\
&&
\alpha(\lambda)\beta(\mu)=\frac{\lambda-\mu-1}{\lambda-\mu}\beta(\mu)\alpha(\lambda)+\frac{1}{\lambda-\mu}\beta(\lambda)\alpha(\mu),
\label{Exch-3}\\
&&
\delta(\lambda)\beta(\mu)=\frac{\lambda-\mu+1}{\lambda-\mu}\beta(\mu)\delta(\lambda)-\frac{1}{\lambda-\mu}\beta(\lambda)\delta(\mu).
\label{Exch-4}
\eea
Let us decompose the double-row monodromy matrix $\mathbb{T}(u)$ in terms of its components which can be expressed in terms
of the components of $T(u)$
\begin{eqnarray}
\mathbb{T}(\lambda)&=&\left(\begin{array}{cc} A(\lambda) & B(\lambda) \\
C(\lambda)& D(\lambda)
  \end{array}\right)\no\\
&=& \left(\begin{array}{cc}
\alpha(\lambda)& \beta{\lambda}\\
\gamma(\lambda) & \delta(\lambda)
\end{array}\right)
\left(\begin{array}{cc}
p+\lambda & 0\\
0 & p-\lambda
\end{array}\right)
\left(\begin{array}{cc}
\delta(-\lambda-1)&-\beta(-\lambda-1) \\
-\gamma(-\lambda-1)&\alpha(-\lambda-1)
\end{array}\right).\no
\end{eqnarray} Namely,
\begin{eqnarray}
&&A(\lambda)=(p+\lambda)\alpha(\lambda)\delta(-\lambda-1)-(p-\lambda)\beta(\lambda)\gamma(-\lambda-1),\nonumber  \\
&&B(\lambda)=-(p+\lambda)\alpha(\lambda)\beta(-\lambda-1)+(p-\lambda)\beta(\lambda)\alpha(-\lambda-1),\label{B-operator} \\
&&C(\lambda)=(p+\lambda)\gamma(\lambda)\delta(-\lambda-1)-(p-\lambda)\delta(\lambda)\gamma(-\lambda-1),\nonumber  \\
&&D(\lambda)=-(p+\lambda)\gamma(\lambda)\beta(-\lambda-1)+(p-\lambda)\delta(\lambda)\alpha(-\lambda-1).\no
\end{eqnarray}
Moreover, the RE (\ref{RE-V}) and the ``RLL" relations (\ref{RLL-1})-(\ref{RLL-3}) enable us to derive exchange relation
of the double-row monodromy matrix $\mathbb{T}(u)$
\bea &&R_{12}(u_1-u_2)\mathbb{T}_1(u_1)R_{21}(u_1+u_2)\mathbb{T}_2(u_2)\no\\
 &&~~~~~~=
\mathbb{T}_2(u_2)R_{12}(u_1+u_2)\mathbb{T}_1(u_1)R_{21}(u_1-u_2),\label{RE-Operator}
\eea which allows one to obtain  the commutation relations among $A(u)$, $B(u)$,
$C(u)$ and $D(u)$.
Here we give the relevant ones given by \cite{Skl88} for our
purpose,
\bea
C(\lambda)B(\mu)&=&B(\mu)C(\lambda) +\frac{\lambda+\mu}{(\lambda+\mu+1)(\lambda-\mu)(2\lambda+1)}A(\mu){\bar D}(\lambda)\nonumber\\
&&+\frac{(\lambda-\mu+1)(2\lambda)}{(\lambda+\mu+1)(\lambda-\mu)(2\lambda+1)}A(\mu)A(\lambda) -\frac{2\lambda[A(\lambda){\bar D}(\mu)+A(\lambda)A(\mu)]}{(\lambda-\mu)(2\lambda+1)(2\mu+1)}\no\\
&&-\frac{[{\bar D}(\lambda){\bar D}(\mu)+{\bar D}(\lambda)A(\mu)]}{(\lambda+\mu+1)(2\lambda+1)(2\mu+1)},\label{BC}\\
A(\lambda)B(\mu)&=&\frac{(\lambda+\mu)(\lambda-\mu-1)}{(\lambda-\mu)(\lambda+\mu+1)}B(\mu)A(\lambda)
-\frac{1}{(\lambda+\mu+1)(2\mu+1)}B(\lambda){\bar D}(\mu)\nonumber\\
&&+\frac{2\mu}{(\lambda-\mu)(2\mu+1)}B(\lambda)A(\mu),\label{AB}\\
{\bar
D}(\lambda)B(\mu)&=&\frac{(\lambda-\mu+1)(\lambda+\mu+2)}{(\lambda-\mu)(\lambda+\mu+1)}B(\mu){\bar
D} (\lambda)-\frac{2(\lambda+1)}{(\lambda-\mu)(2\mu+1)}B(\lambda){\bar D}(\mu)\nonumber\\
&&+\frac{4(\lambda+1)\mu}{(2\mu+1)(\lambda+\mu+1)}B(\lambda)A(\mu).\label{DB}
\eea
Here we have introduced  the operator $\bar D(\lambda)=(2\lambda+1)D(\lambda)-A(\lambda)$. At this stage, we have provided
most of the ingredients required to obtain a fundamental functional relation the eigenvalue of the
transfer matrix (\ref{trans}) with K-matrices given by (\ref{K-}) and (\ref{K+}).


\section{ Functional relations and the T-Q relation}
\label{BAE} \setcounter{equation}{0}

Following the method in \cite{Yan08,Beh96} and using the crossing relation of the
R-matrix (\ref{crosing-unitarity}) and the explicit expressions of the K-matrices
(\ref{K-}) and (\ref{K+}), one can show that the corresponding transfer matrix
$\tau(u)$ has the following
properties,
\bea
&&\hspace{-1.5cm}\mbox{Crossing
symmetry}:\,\quad\tau(-u-1)=\tau(u),\label{T-Cro}\\
&&\hspace{-1.5cm}\mbox{Initial
condition}:\,\tau(0)=2p\,q\prod_{j=1}^N(1-\theta_j)(1+\theta_j)\times\,{\rm id},\label{T-In}\\
&&\hspace{-1.5cm}\mbox{Asymptotic behavior}:\,
\tau(u)\sim 2u^{2N+2}\times \,{\rm id}+\ldots,\quad {\rm for}\, u\rightarrow
\pm\infty.\label{T-Asy} \eea

The commutativity of the transfer matrix $\tau(u)$ implies that one
can find the common eigenstate of $\tau(u)$, which indeed does not
depend upon $u$. Suppose $\langle\Psi|$ is an eigenstate of
$\tau(u)$ with an eigenvalue $\Lambda(u)$, namely, \bea
\langle\Psi|\tau(u) =\Lambda(u)\langle\Psi|.\no \eea The properties
of the transfer matrix $\tau(u)$ given by
(\ref{T-Cro})-(\ref{T-Asy}) imply that the corresponding eigenvalue
$\Lambda(u)$ satisfies the following relations: \bea
&&\hspace{-1.5cm}\mbox{Crossing
symmetry}:\,\quad\L(-u-1)=\L(u),\label{Eig-Cro}\\
&&\hspace{-1.5cm}\mbox{Initial
condition}:\,\L(0)=2p\,q\prod_{j=1}^N(1-\theta_j)(1+\theta_j)=\L(-1),\label{Eig-In}\\
&&\hspace{-1.5cm}\mbox{Asymptotic behavior}:\,
\L(u)\sim 2u^{2N+2}+\ldots,\quad {\rm for}\, u\rightarrow
\pm\infty.\label{Eig-Asy} \eea
The analyticity of the R-matrix and K-matrices and independence on $u$ of the eigenstate
lead to that the eigenvalue $\Lambda(u)$ further obeys  the property
\bea
\hspace{-0.2cm}\mbox{Analyticity}:\,\quad \L(u) \mbox{, as a function of $u$, is a polynomial of degree $2N+2$}.\label{Eigen-Anal}
\eea
Therefore the values of $\L(u)$ at generic $2N+3$ points suffice to determine  the function uniquely. However, we have already
obtained the corresponding values of  $\L(u)$ at points $u=0,\,-1,\,\infty$. For this purpose, we shall find the associated
equations to determine the values of  $\L(u)$ at the other $2N$ points, for example at  $u=\theta_j,-\theta_j-1$.

Let us define the state $|0\rangle=\otimes |\uparrow\rangle_j$ which
now has nothing to do with the eigenstate of the transfer matrix
$\tau(u)$ as it does in the conventional algebraic Bethe ansatz
\cite{Kor93}. On the other hand,  the components of the double-row
monodromy matrix $\mathbb{T}(u)$ act on the states, giving rise to:
\bea
&& C(u)|0\rangle=0, \label{C-1} \\
&&
A(u)|0\rangle=(p+u)\prod_{j=1}^{N}(u-\theta_j+1)(u+\theta_j+1)|0\rangle
=a(u)|0\rangle, \label{a(u)} \\
&& {\bar
D}(u)|0\rangle=2u(p-u-1)\prod_{j=1}^{N}(u-\theta_j)(u+\theta_j)|0\rangle=d(u)|0\rangle.
\label{d(u)} \eea From the explicit expression (\ref{K+}) of the
K-matrix $K^+(u)$, one can express the transfer matrix in terms of
the components of the monodromy matrix $\mathbb{T}(u)$, \bea
\tau(u)&=&(q+u+1)A(u)+\xi(u+1)[C(u)+B(u)]+(q-u-1)D(u)\nonumber\\
&=&
\frac{2(u+q)(u+1)}{2u+1}A(u)+\xi(u+1)[B(u)+C(u)]+\frac{q-u-1}{2u+1}{\bar D}(u).
\eea
Now let us compute the action of $\tau(\theta_j)\tau(\theta_j-1)$ on the state $|0\rangle$.
Keeping in mind  the fact that the function $a(u)$ (or $d(u)$) vanishes at $u=\theta_j-1$ (or $u=\theta_j$), namely,
\bea
a(\theta_j-1)=0=d(\theta_j), \quad\quad j=1,...N,\label{Identity}
\eea
and using the relations (\ref{C-1})-(\ref{d(u)}), we have
\bea
\tau(\theta_j)\tau(\theta_j-1)|0\rangle&=&\frac{2(\theta_j+q)(\theta_j+1)(q-\theta_j)}{(2\theta_j+1)(2\theta_j-1)}
 a(\theta_j)d(\theta_j-1)|0\rangle+\xi^2\theta_j(\theta_j+1)C(\theta_j)B(\theta_j-1)|0\rangle\no\\
&&\hspace{-2.2truecm} +\frac{\xi(q-\theta_j)(\theta_j+1)}{2\theta_j-1}d(\theta_j-1)B(\theta_j)|0\rangle
+\frac{2\xi\theta_j(\theta_j+q)(\theta_j+1)}{2\theta_j+1}A(\theta_j)B(\theta_j-1)|0\rangle\no\\
&&\hspace{-2.2truecm}+\frac{\xi\theta_j(q-\theta_j-1)}{2\theta_j+1}\bar{D}(\theta_j)B(\theta_j-1)|0\rangle
+\xi^2\theta_j(\theta_j+1)B(\theta_j)B(\theta_j-1)|0\rangle\no\\
&&\hspace{-2.6truecm}=\frac{2(\theta_j+1)(q^2-(1+\xi^2)\theta^2_j)}{(2\theta_j-1)(2\theta_j+1)}a(\theta_j)d(\theta_j-1)|0\rangle
+\xi^2\theta_j(\theta_j+1)B(\theta_j)B(\theta_j-1)|0\rangle.\label{BB-T}
\eea In the derivation of the second equality of the above equation, we have used the exchange relation (\ref{BC})-(\ref{DB})
and (\ref{Identity}). One can compute the last terms of (\ref{BB-T}) by expanding the operator $B(u)$ in terms of the components
of one-row monodromy matrix $T(u)$ via (\ref{B-operator}). Direct calculation shows that
\bea
B(\theta_j)B(\theta_j-1)=0.\label{BB-Main}
\eea
The proof of the above very relation is relegated to Appendix A. Finally, we obtain
\bea
\tau(\theta_j)\tau(\theta_j-1)|0\rangle=\frac{2(\theta_j+1)(q^2-(1+\xi^2)\theta^2_j)}{(2\theta_j-1)(2\theta_j+1)}a(\theta_j)d(\theta_j-1)
|0\rangle.
\eea
Multiplying the eigenstate $\langle\Psi|$ from the left on the both sides of the above equation, we have
\bea
 \langle\Psi|\tau(\theta_j)\tau(\theta_j-1)|0\rangle=
 \frac{2(\theta_j+1)(q^2-(1+\xi^2)\theta^2_j)}{(2\theta_j-1)(2\theta_j+1)}a(\theta_j)d(\theta_j-1)
 \langle\Psi|0\rangle,
\eea which leads to the following equations\footnote{Here we assume that the scalar product $\langle\Psi|0\rangle$ is nonzero. The relations (\ref{BB-Main}) and (\ref{Main})
were also obtained previously by the Sklyanin's separation of variables method for the boundary parameters with some constraint \cite{Fra08}.}
\bea
\Lambda(\theta_j)\Lambda(\theta_j-1)&=&\frac{2(\theta_j+1)(q^2-(1+\xi^2)\theta^2_j)}{(2\theta_j-1)(2\theta_j+1)}a(\theta_j)d(\theta_j-1)\no\\
&=& \frac{\Delta_q(\theta_j)}{(1-2\theta_j)(1+2\theta_j)},
\quad\quad j=1,\ldots,N.\label{Main}
\eea The expression of the quantum determinant $\Delta_q(u)$ is given by \cite{Mez90,Zho96}
\bea
\Delta_q(u)={\rm Det}\{T(u)\}\, {\rm Det}\{\hat{T}(u)\}
\,{\rm Det}\{K^-(u)\}\,{\rm Det}\{K^+(u)\},
\eea and the corresponding determinants can be directly calculated
 \bea
\rm{Det}\lt\{T(u)\rt\}\,{\rm
id}&=&tr_{12}\lt(P^{(-)}_{12}T_1(u-1)T_2(u)P^{(-)}_{12}\rt)
=\prod_{j=1}^N(u-\theta_j+1)(u-\theta_j-1)\,{\rm
id},\no\\
\rm{Det}\lt\{\hat{T}(u)\rt\}\,{\rm
id}&=&tr_{12}\lt(P^{(-)}_{12}\hat{T}_1(u-1)\hat{T}_2(u)P^{(-)}_{12}\rt)
=\prod_{j=1}^N(u+\theta_j+1)(u+\theta_j-1)\,{\rm
id},\no\\
\rm{Det}\lt\{K^-(u)\rt\}&=&tr_{12}
\lt(P^{(-)}_{12}K^-_1(u-1)R_{12}(2u-1)K^-_2(u)\rt)=2(u-1)(p^2-u^2),\no\\
\rm{Det}\lt\{K^+(u)\rt\}&=&tr_{12}
\lt(P^{(-)}_{12}K^+_2(u)R_{12}(-2u-1)K^+_1(u-1)\rt)=2(u+1)((1+\xi^2)u^2-q^2).\no
\eea

The equations (\ref{Main}) play an important role in our method,
which gives rise to the functional relations between eigenvalues
$\Lambda(u)$ of the transfer matrix (the trace of the monodromy
matrix algebra) and its quantum determinant $\Delta_q(u)$, i.e.,
$\Lambda(\theta_j)\Lambda(\theta_j-1)\sim\Delta_q(\theta_j)$ when
the spectrum $u$ is taken at some particular points $\{\theta_j\}$
which are the zero points of $B(u)B(u-1)|0\rangle=0$. Since the
trace and the determinant are two basic quantities of a matrix which
are independent of the representation basis, this method can
overcome the obstacle of absence of a reference state which is
crucial in the conventional Bethe ansatz methods. Moreover, the
equations (\ref{Eig-Cro})-(\ref{Eigen-Anal}) and (\ref{Main}) enable
one to determine the functions $\Lambda(u)$. In the following part
of the section, we shall express the solutions of these equations in
terms of a generalized  $T-Q$ relation formulism.

Let us introduce some functions $\bar{a}^{(\pm)}(u)$ and $\bar{d}^{(\pm)}(u)$
\bea
&&\bar{a}^{(\pm)}(u)=\frac{2u+2}{2u+1}(u\pm p)(\sqrt{1+\xi^2}\,u\pm q)\prod_{j=1}^{N}(u+\theta_j+1)(u-\theta_j+1),\label{a-function}\\
&&\bar{d}^{(\pm)}(u)=\frac{2u}{2u+1}(u\mp p+1)(\sqrt{1+\xi^2}\,(u+1)\mp q)
\prod_{j=1}^{N}(u+\theta_j)(u-\theta_j)\no\\
&&\quad\quad\quad =\bar{a}^{(\pm)}(-u-1).\label{d-function}
\eea We can construct the solutions of (\ref{Eig-Cro})-(\ref{Eigen-Anal}) by the following ansatz
\bea
\Lambda^{(\pm)}(u)&=& \bar{a}^{(\pm)}(u)\frac{Q^{(\pm)}(u-1)Q_1^{(\pm)}(u-1)}{Q^{(\pm)}(u)Q_2^{(\pm)}(u)}+
\bar{d}^{(\pm)}(u)\frac{Q^{(\pm)}(u+1)Q_2^{(\pm)}(u+1)}{Q^{(\pm)}(u)Q_1^{(\pm)}(u)}\no\\
&&+2(1\hspace{-0.12truecm}-\hspace{-0.12truecm}\sqrt{1\hspace{-0.08truecm}+\hspace{-0.08truecm}\xi^2})
u(u\hspace{-0.12truecm}+\hspace{-0.12truecm}1)\frac{\prod_{j=1}^N
(u\hspace{-0.08truecm}+\hspace{-0.08truecm}\theta_j)(u\hspace{-0.08truecm}-\hspace{-0.08truecm}\theta_j)
(u\hspace{-0.08truecm}+\hspace{-0.08truecm}\theta_j\hspace{-0.08truecm}+\hspace{-0.08truecm}1)
(u\hspace{-0.08truecm}-\hspace{-0.08truecm}\theta_j\hspace{-0.08truecm}+\hspace{-0.08truecm}1)}
{Q^{(\pm)}(u)Q_1^{(\pm)}(u)Q_2^{(\pm)}(u)}.
\label{T-Q}
\eea The functions $Q^{(\pm)}(u)$, $Q_1^{(\pm)}(u)$ and $Q_2^{(\pm)}(u)$ are parameterized by $N$  Bethe roots
$\{\lambda^{(\pm)}_j|j=1,\ldots N-2M\}$, $\{\mu^{(\pm)}_j|j=1,\ldots,M\}$ and $\{\nu^{(\pm)}_j|j=1,\ldots,M\}$ with
$M=0,\ldots,\lt[\frac{N}{2}\rt]$ as follows,
\bea
 Q^{(\pm)}(u)&=&\prod_{j=1}^{N-2M}(u-\lambda^{(\pm)}_j)(u+\lambda^{(\pm)}_j+1),\label{Q-1}\\
 Q_1^{(\pm)}(u)&=&\prod_{j=1}^{M}(u-\mu^{(\pm)}_j)(u+\nu^{(\pm)}_j+1),\label{Q-2}\\
 Q_2^{(\pm)}(u)&=&\prod_{j=1}^{M}(u-\nu^{(\pm)}_j)(u+\mu^{(\pm)}_j+1).\label{Q-3}
\eea These $N$ parameters are different from each other and  satisfy  the following Bethe ansatz equations
\bea
&&-\frac{\bar{a}^{(\pm)}(\lambda^{(\pm)}_j)} {\bar{d}^{(\pm)}(\lambda^{(\pm)}_j)}
-\frac{Q^{(\pm)}(\lambda^{(\pm)}_j+1)Q_2^{(\pm)}(\lambda^{(\pm)}_j)Q_2^{(\pm)}(\lambda^{(\pm)}_j+1)}
{Q^{(\pm)}(\lambda^{(\pm)}_j-1)Q_1^{(\pm)}(\lambda^{(\pm)}_j)Q_1^{(\pm)}(\lambda^{(\pm)}_j-1)}\no\\
 &&\quad\quad=\frac{(1\hspace{-0.08truecm}-\hspace{-0.08truecm}\sqrt{1\hspace{-0.08truecm}+\hspace{-0.08truecm}\xi^2})
 (\lambda^{(\pm)}_j\hspace{-0.08truecm}+\hspace{-0.08truecm}1)(2\lambda^{(\pm)}_j\hspace{-0.08truecm}+\hspace{-0.08truecm}1)
 \prod_{l=1}(\lambda^{(\pm)}_j\hspace{-0.08truecm}+\hspace{-0.08truecm}\theta_l\hspace{-0.08truecm}+\hspace{-0.08truecm}1)
(\lambda^{(\pm)}_j\hspace{-0.08truecm}-\hspace{-0.08truecm}\theta_l\hspace{-0.08truecm}+\hspace{-0.08truecm}1)}
{(\lambda^{(\pm)}_j\mp p+1)(\sqrt{1+\xi^2}(\lambda^{(\pm)}_j+1)\mp q)
Q^{(\pm)}(\lambda^{(\pm)}_j-1)Q_1^{(\pm)}(\lambda^{(\pm)}_j)Q_1^{(\pm)}(\lambda^{(\pm)}_j-1)},\no\\
&&\quad\quad j=1,\ldots,N-2M,\label{BAE-M-1}\\
&&\frac{(1\hspace{-0.08truecm}-\hspace{-0.08truecm}\sqrt{1\hspace{-0.08truecm}+\hspace{-0.08truecm}\xi^2})
 (\mu^{(\pm)}_j\hspace{-0.08truecm}+\hspace{-0.08truecm}1)(2\mu^{(\pm)}_j\hspace{-0.08truecm}+\hspace{-0.08truecm}1)
 \prod_{l=1}(\mu^{(\pm)}_j\hspace{-0.08truecm}+\hspace{-0.08truecm}\theta_l\hspace{-0.08truecm}+\hspace{-0.08truecm}1)
(\mu^{(\pm)}_j\hspace{-0.08truecm}-\hspace{-0.08truecm}\theta_l\hspace{-0.08truecm}+\hspace{-0.08truecm}1)}
{(\mu^{(\pm)}_j\mp p+1)(\sqrt{1+\xi^2}(\mu^{(\pm)}_j+1)\mp q)
Q^{(\pm)}(\mu^{(\pm)}_j+1)Q_2^{(\pm)}(\mu^{(\pm)}_j)Q_2^{(\pm)}(\mu^{(\pm)}_j+1)}=-1,\no\\
&&\quad\quad j=1,\ldots,M,\label{BAE-M-2}\\
&&\frac{(1\hspace{-0.08truecm}-\hspace{-0.08truecm}\sqrt{1\hspace{-0.08truecm}+\hspace{-0.08truecm}\xi^2})
 \nu^{(\pm)}_j(2\nu^{(\pm)}_j\hspace{-0.08truecm}+\hspace{-0.08truecm}1)
 \prod_{l=1}(\nu^{(\pm)}_j\hspace{-0.08truecm}+\hspace{-0.08truecm}\theta_l)
(\nu^{(\pm)}_j\hspace{-0.08truecm}-\hspace{-0.08truecm}\theta_l)}
{(\nu^{(\pm)}_j\pm p)(\sqrt{1+\xi^2}\nu^{(\pm)}_j\pm q)
Q^{(\pm)}(\nu^{(\pm)}_j-1)Q_1^{(\pm)}(\nu^{(\pm)}_j)Q_1^{(\pm)}(\nu^{(\pm)}_j-1)}=-1,\no\\
&&\quad\quad j=1,\ldots,M.\label{BAE-M-3}
\eea

Some remarks are in order.  The completeness \cite{Nep03,Yan06} of
the two sets of  eigenvalues in the XXZ spin chain with constrained
unparallel boundary fields case suggests that in our case, the
eigenvalues $\Lambda^{(-)}(u)$ and $\Lambda^{(+)}(u)$ {\it together}
constitute the complete set of eigenvalues of the transfer matrix
$t(u)$ of the open XXX spin chain. The last term of our generalized
$T-Q$ relation (\ref{T-Q}) (c.f. the conventional type \cite{Bax82}
) is crucial, which has encoded the contribution from the
off-diagonal element of the K-matrix and vanishes when two boundary
fields are parallel. Taking the limit: $\xi\rightarrow 0$, the set
of parameter $\{\mu_j\}$ coincides with that of $\{\nu_j\}$, the
corresponding $T-Q$ relation (\ref{T-Q}) becomes the usual type \bea
\Lambda^{(\pm)}(u)&=&
\bar{a}^{(\pm)}(u)\frac{\bar{Q}^{(\pm)}(u-1)}{\bar{Q}^{(\pm)}(u)}+
\bar{d}^{(\pm)}(u)\frac{\bar{Q}^{(\pm)}(u+1)}{\bar{Q}^{(\pm)}(u)},\label{T-Q-Ordinary}
\eea where $\bar{Q}^{(\pm)}(u)=Q^{(\pm)}(u)Q_1^{(\pm)}(u)$. Then the
resulting solutions are reduced to those in \cite{Skl88}.


\section{Homogeneous limit}
\label{Hom} \setcounter{equation}{0}

The results of the previous sections focus on  the inhomogeneous spin-$\frac{1}{2}$ XXX open chain.
In this section we consider the homogeneous limit,  i.e. $\{\theta_j\}\longrightarrow 0$, of the above results.
The corresponding
eigenvalue $\Lambda(u)$ of the transfer matrix of the homogeneous model satisfies the following relations:
\bea
&&\hspace{-1.5cm}\mbox{Crossing
symmetry}:\,\quad\L(-u-1)=\L(u),\label{H-Eig-Cro}\\
&&\hspace{-1.5cm}\mbox{Initial
condition}:\,\L(0)=2p\,q=\L(-1),\label{H-Eig-In}\\
&&\hspace{-1.5cm}\mbox{Asymptotic behavior}:\,
\L(u)\sim 2u^{2N+2}+\ldots,\quad {\rm for}\, u\rightarrow
\pm\infty,\label{H-Eig-Asy}\\
&&\hspace{-1.5cm}\mbox{Analyticity}:\,\quad \L(u) \mbox{, as a
function of $u$, is a polynomial of degree
$2N+2$}.\label{H-Eigen-Anal} \eea The corresponding relations
(\ref{Main}) now read \bea \frac{\partial^{l}}{\partial
u^l}\lt\{\Lambda(u)\Lambda(u-1)\rt\}|_{u=0}=
\frac{\partial^{l}}{\partial
u^l}\lt\{\frac{\bar{\Delta}_q(u)}{(1+2u)(1-2u)}\rt\}|_{u=0},\,\,l=0,1,\ldots,2N-1.\label{Main-H}
\eea Here the quantum determinant $\bar{\Delta}_q(u)$ becomes \bea
\bar{\Delta}_q(u)=4(u^2-1)(p^2-u^2)((1+\xi^2)^2u^2-q^2)(u+1)^{2N}(u-1)^{2N}.
\eea Actually the equations (\ref{H-Eig-Cro})-(\ref{Main-H}) enable
one to determine the functions $\Lambda(u)$. We shall express the
solutions of these equations in terms of a generalized  $T-Q$
relation formulism. For this purpose, let us introduce the following
functions $H^{(\pm)}(u)$ and $\Lambda^{(\pm)}(u)$ \bea
H^{(\pm)}(u)&=&\frac{2u+2}{2u+1}(u\pm p)(\sqrt{1+\xi^2}\,u\pm q)(u+1)^{2N},\label{H-function}\\
\Lambda^{(\pm)}(u)&=& H^{(\pm)}(u)\frac{Q^{(\pm)}(u-1)Q_1^{(\pm)}(u-1)}{Q^{(\pm)}(u)Q_2^{(\pm)}(u)}+
H^{(\pm)}(-u-1)\frac{Q^{(\pm)}(u+1)Q_2^{(\pm)}(u+1)}{Q^{(\pm)}(u)Q_1^{(\pm)}(u)}\no\\
&&\quad\quad+\frac{2(1-\sqrt{1+\xi^2})
u^{2N+1}(u+1)^{2N+1}}{Q^{(\pm)}(u)Q_1^{(\pm)}(u)Q_2^{(\pm)}(u)},
\label{H-T-Q}
\eea where the functions $Q^{(\pm)}(u)$, $Q_1^{(\pm)}(u)$ and $Q_2^{(\pm)}(u)$ are also given by (\ref{Q-1})-(\ref{Q-3}).
Then the functions $\Lambda^{(\pm)}(u)$ become the solutions of
the equations (\ref{H-Eig-Cro})-(\ref{Main-H}) provided that the parameters
$\{\lambda^{(\pm)}_{j}\}$, $\{\mu^{(\pm)}_{j}\}$  and $\{\nu^{(\pm)}_{j}\}$ satisfy the following Bethe ansatz equations respectively
\bea
&&-\frac{H^{(\pm)}(\lambda^{(\pm)}_j)} {H^{(\pm)}(-\lambda^{(\pm)}_j-1)}-\frac{Q^{(\pm)}(\lambda^{(\pm)}_j+1)Q^{(\pm)}_2(\lambda^{(\pm)}_j)Q^{(\pm)}_2(\lambda^{(\pm)}_j+1)}
{Q^{(\pm)}(\lambda^{(\pm)}_j-1)Q^{(\pm)}_1(\lambda^{(\pm)}_j)Q^{(\pm)}_1(\lambda^{(\pm)}_j-1)}\no\\
 &&\quad\quad=\frac{(1\hspace{-0.08truecm}-\hspace{-0.08truecm}\sqrt{1\hspace{-0.08truecm}+\hspace{-0.08truecm}\xi^2})
 (\lambda^{(\pm)}_j\hspace{-0.08truecm}+\hspace{-0.08truecm}1)(2\lambda^{(\pm)}_j\hspace{-0.08truecm}+\hspace{-0.08truecm}1)
(\lambda^{(\pm)}_j\hspace{-0.08truecm}+\hspace{-0.08truecm}1)^{2N}}
{(\lambda^{(\pm)}_j\mp p+1)(\sqrt{1+\xi^2}(\lambda^{(\pm)}_j+1)\mp q)Q^{(\pm)}(\lambda^{(\pm)}_j\hspace{-0.08truecm}-\hspace{-0.08truecm}1)
Q^{(\pm)}_1(\lambda^{(\pm)}_j)Q^{(\pm)}_1(\lambda^{(\pm)}_j\hspace{-0.08truecm}-\hspace{-0.08truecm}1)},\no\\
&&\quad\quad\quad\quad\quad\quad j=1,\ldots,N-2M,\label{BAE-M-H-1}\\
&&\frac{(1\hspace{-0.08truecm}-\hspace{-0.08truecm}\sqrt{1\hspace{-0.08truecm}+\hspace{-0.08truecm}\xi^2})
 (\mu^{(\pm)}_j\hspace{-0.08truecm}+\hspace{-0.08truecm}1)(2\mu^{(\pm)}_j\hspace{-0.08truecm}+\hspace{-0.08truecm}1)
(\mu^{(\pm)}_j\hspace{-0.08truecm}+\hspace{-0.08truecm}1)^{2N}}
{(\mu^{(\pm)}_j\mp p+1)(\sqrt{1+\xi^2}(\mu^{(\pm)}_j+1)\mp q)
Q^{(\pm)}(\mu^{(\pm)}_j+1)Q_2^{(\pm)}(\mu^{(\pm)}_j)Q_2^{(\pm)}(\mu^{(\pm)}_j+1)}=-1,\no\\
&&\quad\quad\quad\quad\quad\quad j=1,\ldots,M,\label{BAE-M-H-2}\\
&&\frac{(1\hspace{-0.08truecm}-\hspace{-0.08truecm}\sqrt{1\hspace{-0.08truecm}+\hspace{-0.08truecm}\xi^2})
 \nu^{(\pm)}_j(2\nu^{(\pm)}_j\hspace{-0.08truecm}+\hspace{-0.08truecm}1)
 {\nu^{(\pm)}_j}^{2N}}
{(\nu^{(\pm)}_j\pm p)(\sqrt{1+\xi^2}\nu^{(\pm)}_j\pm q)
Q^{(\pm)}(\nu^{(\pm)}_j-1)Q_1^{(\pm)}(\nu^{(\pm)}_j)Q_1^{(\pm)}(\nu^{(\pm)}_j-1)}=-1,\no\\
&&\quad\quad\quad\quad\quad\quad j=1,\ldots,M.\label{BAE-M-H-3}
\eea

With the help of the relation (\ref{oh}) between the Hamiltonian
(\ref{ohami}) and the transfer matrix $\tau(u)$, the energy value
$E$ of the Hamiltonian is given as \bea
&&E^{(\pm)}=2\sum_{j=1}^{N-2M}\frac{1}{\lambda^{(\pm)}_j(\lambda^{(\pm)}_j+1)}
+2\sum_{j=1}^{M}\lt(\frac{1}{\nu^{(\pm)}_j}-\frac{1}{\mu^{(\pm)}_j+1}\rt)
\no \\
&&\qquad \quad
+N-1+\frac1p+\frac{\sqrt{1+\xi^2}}{q},\label{Spectrum} \eea where
the parameters $\{\lambda^{(\pm)}_{j}\}$ and $\{\mu^{(\pm)}_{j}\}$
need to satisfy the associated  Bethe ansatz equations
(\ref{BAE-M-H-1})-(\ref{BAE-M-H-3}) respectively. It would be
believed that the eigenvalues $E^{(-)}(u)$ and $E^{(+)}(u)$ {\it
together} constitute the complete set of energy spectrum of the
Hamiltonian (\ref{ohami}).


\section{Discussions}
\label{Con} \setcounter{equation}{0}

In this paper, the open spin-$\frac{1}{2}$ XXX spin chain with
unparallel boundary fields defined by the Hamiltonian (\ref{ohami})
is exactly diagonalized by the off-diagonal Bethe ansatz method
proposed in \cite{Cao13}. There are two sets of eigenvalues
$E^{(\pm)}$ given by (\ref{Spectrum}), which are expressed in terms
of the roots of the associated Bethe ansatz equations
(\ref{BAE-M-H-1})-(\ref{BAE-M-H-3}).

As for integrable models without $U(1)$ symmetry, some off-diagonal
elements of monodromy matrix enter into expression of the transfer
matrix. This breaks down the usual $U(1)$ symmetry. However the
central idea of our method is to construct the functional relations
such as (\ref{Main}) between eigenvalues $\Lambda(\lambda)$ of the
transfer matrix (the trace of the monodromy matrix) and its quantum
determinant $\Delta_q(\lambda)$ based on the zero points of the
product of off-diagonal elements of monodromy matrix
$B(u)B(u-\eta)=0$. In fact, with some operator product identities,
we can demonstrate \footnote{We have directly proven such an operator identity in \cite{Cao13-1}.}
\begin{eqnarray}
\tau(\theta_j)\tau(\theta_j-1)=
\frac{\Delta_q(\theta_j)}{(1-2\theta_j)(1+2\theta_j)}, \quad\quad
j=1,\ldots,N,\label{Main2}
\end{eqnarray}
which is completely independent of the representation basis and thus
avoids the obstacle of absence of a reference state which is crucial
in the conventional Bethe ansatz methods. Although the functional
relations between eigenvalues $\Lambda(\lambda)$  and the quantum
determinant $\Delta_q(\lambda)$ at some particular points can be
obtained by different ways \cite{Cao13,Nic12,Fra11}, the very
relations would play an important role in the method. In our case,
based on the relation (\ref{Main}) and some properties
(\ref{Eig-Cro})-(\ref{Eigen-Anal}) of $\Lambda(\lambda)$ we can give
a generalized $T-Q$ relation type solution (\ref{T-Q}), which
modifies the usual $T-Q$ relation by adding an extra term. Such an
extra term encodes the contribution of the off-diagonal element of
the associated K-matrix.

The numerical results\cite{Nep03} strongly suggest that a fixed $M$
might give a complete set of eigenvalues of the transfer matrix. In such a sense,
different $M$ might only give different parametrization of the eigenvalues but not different
states. This allows us to simplify further the generalized $T-Q$ relation as follows.
For the case of $N$ being even, let $M=\frac{N}{2}$ and the corresponding
eigenvalue $\Lambda(u)$ can be parameterized by
\bea
\Lambda^{(\pm)}(u)&=& \bar{a}^{(\pm)}(u)\frac{Q_1^{(\pm)}(u-1)}{Q_2^{(\pm)}(u)}+
\bar{d}^{(\pm)}(u)\frac{Q_2^{(\pm)}(u+1)}{Q_1^{(\pm)}(u)}\no\\
&&+2(1\hspace{-0.12truecm}-\hspace{-0.12truecm}\sqrt{1\hspace{-0.08truecm}+\hspace{-0.08truecm}\xi^2})
u(u\hspace{-0.12truecm}+\hspace{-0.12truecm}1)\frac{\prod_{j=1}^N
(u\hspace{-0.08truecm}+\hspace{-0.08truecm}\theta_j)(u\hspace{-0.08truecm}-\hspace{-0.08truecm}\theta_j)
(u\hspace{-0.08truecm}+\hspace{-0.08truecm}\theta_j\hspace{-0.08truecm}+\hspace{-0.08truecm}1)
(u\hspace{-0.08truecm}-\hspace{-0.08truecm}\theta_j\hspace{-0.08truecm}+\hspace{-0.08truecm}1)}
{Q_1^{(\pm)}(u)Q_2^{(\pm)}(u)}. \label{T-Q-1} \eea For the case of
$N$ being odd, let $M=\frac{N+1}{2}$ and the corresponding
eigenvalue $\Lambda(u)$ can be parameterized by \bea
\Lambda^{(\pm)}(u)&=&
\bar{a}^{(\pm)}(u)\frac{Q_1^{(\pm)}(u-1)}{Q_2^{(\pm)}(u)}+
\bar{d}^{(\pm)}(u)\frac{Q_2^{(\pm)}(u+1)}{Q_1^{(\pm)}(u)}\no\\
&&+2(1\hspace{-0.12truecm}-\hspace{-0.12truecm}\sqrt{1\hspace{-0.08truecm}+\hspace{-0.08truecm}\xi^2})
u^2(u\hspace{-0.12truecm}+\hspace{-0.12truecm}1)^2\frac{\prod_{j=1}^N
(u\hspace{-0.08truecm}+\hspace{-0.08truecm}\theta_j)(u\hspace{-0.08truecm}-\hspace{-0.08truecm}\theta_j)
(u\hspace{-0.08truecm}+\hspace{-0.08truecm}\theta_j\hspace{-0.08truecm}+\hspace{-0.08truecm}1)
(u\hspace{-0.08truecm}-\hspace{-0.08truecm}\theta_j\hspace{-0.08truecm}+\hspace{-0.08truecm}1)}
{Q_1^{(\pm)}(u)Q_2^{(\pm)}(u)}. \label{T-Q-2} \eea The above $T-Q$
relations lead to the Bethe ansatz equations which allow one to
investigate the distribution of roots of these equations and compute
the physical properties in the thermodynamic limit by the usual
method \cite{Kor93}.

\section*{Acknowledgments}

The financial support from  the National Natural Science Foundation
of China (Grant Nos. 11174335, 11075126 and 11031005), the National
Program for Basic Research of MOST (973 project under grant
No.2011CB921700) and the State Education Ministry of China (Grant
No. 20116101110017 and SRF for ROCS) are gratefully acknowledged.
Two of the authors (W. Yang and K. Shi) would like to thank IoP/CAS
for the hospitality and they enjoyed during their visit there.


\section*{Appendix A: Proof of (\ref{BB-Main}) }
\setcounter{equation}{0}
\renewcommand{\theequation}{A.\arabic{equation}}

In this appendix, we prove the relation (\ref{BB-Main}). From the definition (\ref{Mon-V-1}) of the
one-row monodromy matrix $T(u)$  and the exchange relations (\ref{Exch-1})-(\ref{Exch-4})
of its components, we have
\begin{eqnarray}
&& \alpha(\theta_j)\beta(\theta_j-1)|0\rangle=0=\delta(\theta_j)\alpha(\theta_j-1)|0\rangle, \label{usr1} \\
&& \alpha(\theta_j)\alpha(\theta_j-1)|0\rangle=0=\delta(\theta_j)\delta(\theta_j-1)|0\rangle,\label{usr2}\\
&& \gamma(\theta_j)\beta(\theta_j-1)|0\rangle=-\tilde a(\theta_j)\tilde d(\theta_j-1)|0\rangle, \label{usr3}\\
&& \delta(\theta_j)\beta(\theta_j-1)|0\rangle=-\tilde d(\theta_j-1)\b(\theta_j)|0\rangle, \label{usr4} \\
&& \alpha(\theta_j)\delta(\theta_j-1)|0\rangle=\tilde a(\theta_j)\tilde d(\theta_j-1)|0\rangle, \label{usr5} \\
&& \beta(\theta_j)\beta(\theta_j-1)|0\rangle=0,  \label{usr6}
\end{eqnarray}
where
\begin{eqnarray}
\tilde a(\lambda)=\prod_{j=1}^N(\lambda-\theta_j+1), \quad \tilde
d(\lambda)=\prod_{j=1}^N(\lambda-\theta_j).
\end{eqnarray}
One can check the last equation (\ref{usr6}) by induction. Expanding the operator $B(u)$ in terms of the components of the
one-row monodromy matrix (\ref{B-operator}) and using the relations (\ref{usr1})-(\ref{usr6}), we have
\begin{eqnarray}
B(\theta_j)B(\theta_j-1)|0\rangle&=&\lt\{-(p+\theta_j)\alpha(\theta_j)\beta(-\theta_j-1)
+(p-\theta_j)\beta(\theta_j)\alpha(-\theta_j-1)\rt\}\nonumber \\
&&\times\lt\{-(p+\theta_j-1)\alpha(\theta_j-1)\beta(-\theta_j)+(p-\theta_j+1)\beta(\theta_j-1)\alpha(-\theta_j)\rt\}|0\rangle\nonumber \\
&=&\frac{(2\theta_j+1)(\theta_j-p+1)(2p+(\theta_j-p-1)2\theta_j)}{(\theta_j+1)(2\theta_j-1)}\nonumber \\
&& \times \tilde a(-\theta_j)
\tilde a(-\theta_j-1)\beta(\theta_j)\beta(\theta_j-1)|0\rangle\nonumber \\
&=&0.
\end{eqnarray}
Due to the fact that for a generic value of  $\{u_j\}$ the set of vectors $\lt\{\prod_{j=1}^M B(u_j)|0\rangle|,M=0,\ldots,N\rt\}$ spans the whole
vector space ${\rm\bf V}^{\otimes N}$ and the fact that the commutativity between the operators $B$ with different spectrum, one can conclude that
$B(\theta_j)B(\theta_j-1)=0$. This completes the proof of (\ref{BB-Main})\footnote{One may also check that the operator $B(u)B(u-1)$,
when $u=\theta_j$, becomes zero by using the very properties (\ref{Int-R}) and (\ref{Ant}) of the R-matrix and QYBE and RE relations.}.


\end{document}